# Spontaneous emission control in high-extraction efficiency plasmonic crystals

**Hideo Iwase, Dirk Englund, and Jelena Vučković**

*Ginzton Laboratory, Stanford University, California, 94305, USA*

*Hideo.Iwase@stanford.edu*

*http://ee.stanford.edu/~jela/*

**Abstract:** We experimentally and theoretically investigate exciton-field coupling for the surface plasmon polariton (SPP) in waveguide-confined (WC) anti-symmetric modes of hexagonal plasmonic crystals in InP-TiO-Au-TiO-Si heterostructures. The radiative decay time of the InP-based transverse magnetic (TM)-strained multi-quantum well (MQW) coupled to the SPP modes is observed to be 2.9-3.7 times shorter than that of a bare MQW wafer. Theoretically we find that 80 % of the enhanced PL is emitted into SPP modes, and 17 % of the enhanced luminescence is redirected into WC-anti-symmetric modes. In addition to the direct coupling of the excitons to the plasmonic modes, this demonstration is also useful for the development of high-temperature SPP lasers, the development of highly integrated photo-electrical devices, or miniaturized biosensors.

---

## References and links

1. Y. Gong, and J. Vučković, " Design of plasmon cavities for solid-state cavity quantum electrodynamics applications," Appl. Phys. Lett. **90**, 033113 (2007).
2. K. Okamoto, I. Niki, A. Shvartser, Y. Narukawa, T. Mukai and A. Scherer, "Surface-plasmon-enhanced light emitters based on InGaN quantum wells," Nat. Mater. **3**, 601-605 (2004).
3. A. Neogi, C. Lee, H. O. Everitt, T. Kuroda, A. Tackeuchi, and E. Yablonovitch, "Enhancement of spontaneous recombination rate in a quantum well by resonant surface plasmon coupling," Phys. Rev. B **66**, 153305 (2002).
4. H. Raether, *Surface Plasmons on Smooth and Rough Surfaces and on Gratings* (Springer, Berlin, 1988).
5. D. Sarid, "Long-Range Surface Plasmon Waves on Very Thin Metal Films," Phys. Rev. Lett. **47**, 1927-1930 (1981).
6. M. Hochberg, T. Baehr-Jones, C. Walker, and A. Scherer, "Integrated plasmon and dielectric waveguides," Opt. Express **12**, 5481-5481 (2004).
7. F. Liu, Y. Rao, Y. Huang, W. Zhang, and J. Peng, "Coupling between long range surface plasmon polariton mode and dielectric waveguide mode," Appl. Phys. Lett. **90**, 141101 (2007).
8. T. Nikolajsen, K. Leosson, and S. I. Bozhevolnyi, "Surface Plasmon polariton based modulators and switches operating at telecom wavelengths," Appl. Phys. Lett. **85**, 5833-5835 (2005).
9. I. Gontijo, M. Boroditsky, E. Yablonovitch, S. keller, U. K. Mishra, and S. P. DenBaars, "Coupling of InGaN quantum-well photoluminescence to silver surface plasmons," Phys. Rev. B **60**, 11564-11567 (1999).
10. E. N. Economou, "Surface Plasmons in thin Films," Phys. Rev. **182**, 539-554 (1969).
11. C. Sirtori, C. Gmachl, F. Capasso, J. Faist, D. L. Sivco, A. L. Hutchinson, and A. Y. Cho, "Long-wavelength (λ ≈ 8–11.5 μm) semiconductor lasers with waveguides based on surface plasmons," Opt. Lett. **23**, 1366-1368 (1998).
12. T. Okamoto, F. H'Dhili, and S. Kawata, "Towards plasmonic band gap laser," Appl. Phys. Lett. **85**, 3968-3970 (2004).
13. S. Kumar, B. S. Williams, Q. Qin, A. W. Lee, Q. Hu, and J. L. Reno, "Surface-emitting distributed feedback Terahertz quantum-cascade lasers in metal-metal waveguides," Opt. Express **15**, 113-123 (2007).
14. L. Landau, and E. Lifshitz, *Electrodynamics of Continuum Media* (Pergamon, New York, 1984).

15. J. Vučković, M. Lončar, and A. Scherer, "Surface Plasmon Enhanced Light-Emitting Diode," IEEE J. Quantum Electron. **36**, 1131-1144 (2000).
16. S. C. Kitson, W. L. Barnes, and J. R. Sambles, "Full Photonic Band Gap for Surface Modes in the Visible," Phys. Rev. Lett. **77**, 2670-2673 (1996).
17. L. A. Coldren, and S. W. Corzine, *Diode Lasers and Photonic Integrated Circuits* (Wiley, New York, 1995).
18. R. K. Lee, Y. Xu, and A. Yariv, "Modified spontaneous emission from a two-dimensional photonic bandgap crystal slab," J. Opt. Soc. Am. B**17**, 1438-1442 (2000).
19. S. Adachi, *Physical Properties of III-V semiconductor compounds* (Wiley, New York, 1992).
20. H. F. Ghaemi T. Thio, D. E. Grupp, T. W. Ebbesen, and H. J. Lezec, "Surface plasmons enhance optical transmission through subwavelength holes," Phys. Rev. B**58**, 6779-6782 (1998).

## 1. Introduction

Surface-bound optical excitations with extremely small mode volume can dramatically alter the spontaneous emission rate of nearby emitters [1-3]. In symmetric plasmonic structures, in which the metal membrane is sandwiched by the same dielectric media on its top and bottom, surface-bound optical excitations propagating on two sides of the metal layer are coupled in such a way to produce a field maximum or minimum at the center of the metal layer [4,5]. Among such surface plasmon polaritons (SPPs), modes with odd symmetry across the metal layer reduce ohmic losses. Their field confinement is weaker compared with tightly-confined symmetric modes, but still bounded to the interface, the anti-symmetric modes with far lower ohmic loss hold promise of new plasmonic technologies for biochemical sensing [6], nanocouplers [7], and photonics/electronics intermediaries [8]. Unfortunately, stringent index-matching requirements have prevented these advantages in semiconductor active devices at telecommunication wavelengths [9]. In semiconductor active devices at the telecommunication wavelengths, it has proven extremely difficult to reliably create the required top-bottom index matching. If the matching is not within ~ 1 % for such devices, then the anti-symmetric SPP mode is shifted above the light line, resulting in large leakage losses [9,10]. Hence, as opposed to earlier work which relied on the SPP resonance in the visible [2,3,9], a novel design to achieve Purcell enhancement for SPP modes in the infrared (IR) wavelength range is important for the development of room-temperature plasmonic lasers [11-13] and high-efficiency light emitting diodes, operating at telecommunication wavelength.

## 2. Structure design

We overcome the index matching problem in the infrared by replacing the top layer of a symmetric InP-Au-InP structure with a slightly higher-index silicon layer. The resulting unpatterned underlying InP-TiO-Au-TiO-Si structure is shown in the inset of Fig.1(a), where TiO layers are added as protection against interdiffusion. The corresponding dispersion relations of SPP modes obtained are shown in Fig. 1(a). The solid lines show the dispersion as derived analytically from Maxwell's equations, ignoring metal absorption; we verified these by Finite Difference Time Domain (FDTD) simulations, shown in the dotted line. The anti-symmetric-like SPP modes appear above the light line for the top-layer (Si) and below the light line for the bottom-layer (InP) because of their index mismatch ($n_{Si}$ = 3.4 and $n_{InP}$ = 3.2), but are confined by the waveguide consisting of a Si layer (Fig.1(b)). We refer to these modes as the waveguide-confined (WC) anti-symmetric modes. The symmetric branch is shifted toward lower frequencies in the structure with Si because of SPP overlap with a Si layer [10]. Figure 1(c) shows the distributions of photon energy densities for symmetric and WC-anti-symmetric modes in the unpatterned InP-TiO-Au-TiO-Si structure. The photon energy density is calculated from $\partial(\varepsilon\omega)/\partial\omega \cdot |\mathbf{E}|^2 / 8\pi$ [9,14]. The photon density for the WC-anti-symmetric mode is asymmetric relative to the middle of the Au layer, but still has the minimum inside of the Au layer, thereby leading to low absorption losses. As a result, this structure greatly relaxes the stringent index-matching requirement, while preserving the mode confinement and low leakage losses, and provides low material losses comparable to those in the completely symmetric SPP structures. The mode confinement within ultra-short distance from metal/dielectric interface, well below the optical diffraction limit, provides large Purcell-enhancement for excitons in a nearby quantum well stack.

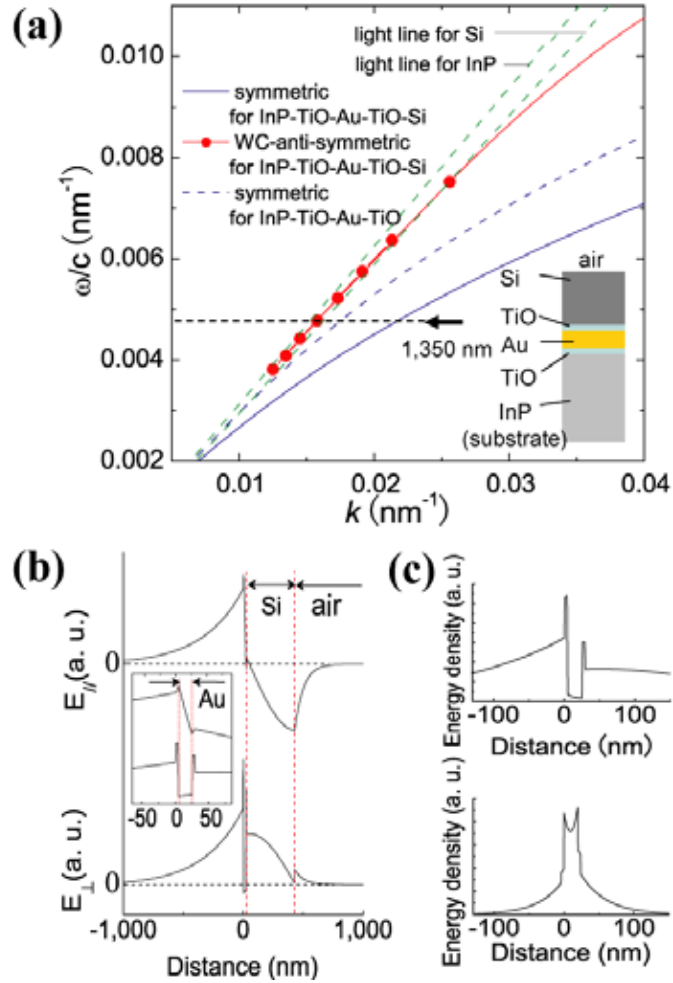

Fig. 1. Dispersion diagrams, field patterns, and distributions of photon energy densities in the proposed metal-dielectric heterostructure. a) Dispersion diagrams for the unpatterned InP-TiO-Au-TiO-Si structure (inset) ,and for an InP-TiO-Au-TiO structure, obtained by analytically solving Maxwell equations (lines) and verified by FDTD simulation (dots). The thicknesses of layers in the inset are 4 nm, 20 nm, 4 nm, and 400 nm respectively from the bottom. b) Parallel ($E_{//}$) and perpendicular ($E_\perp$) electric field components for WC-anti-symmetric SPP at 1,350 nm. The inset shows magnified images of $E_{//}$ (top plot) and $E_\perp$ (bottom plot) around the Au layer. c) Distributions of the photon energy densities for symmetric (bottom) and WC-anti-symmetric modes (top) for InP-TiO-Au-TiO-Si at 1,350 nm. The estimated quality factors due to the metal absorption are around 140 for a WC-anti-symmetric mode and 18 for a symmetric mode at 1,350 nm.

## 3. Experiment

*3.1 Sample preparation*

In our experiment, the InP-TiO-Au-TiO-Si structure is patterned with a hexagonal array to form a plasmonic crystal. Figure 2(a) shows scanning electron microscope (SEM) images of the cross-section of InP-TiO-Au-TiO-Si heterostructure and the fabricated hexagonal plasmonic crystal. The introduction of plasmonic crystals into the heterostructure shown in Fig. 2(a) folds the branches of the dispersion diagram shown in Fig. 1(a) back into the first Brillouin zone and opens mini-bandgaps at high symmetry points [15,16]. This approach redirects the emission coupled to the SPP modes into the vertical direction ($\Gamma$), which is collected in our experiment. For comparison, Fig. 2(b) shows the lateral field patterns for anti-symmetric SPP modes in a similar plasmonic crystal with a perfectly matched index across the metal membrane (InP-Au-InP).

The active region shown in Fig. 2(a) consists of five 0.9 % tensile-strained InP-based multi-quantum wells (MQWs) located 20 nm apart from the TiO layer. The Au, Si, and 2nm-thick-Ti layers were deposited by evaporation. Ti layers were subsequently oxidized in oxygen plasma, to produce 4 nm thick TiO layers that prevent Au migration into Si and InP during the fabrication and the measurement. We confirmed that this oxidation process does not affect the spectrum of MQWs relative to its intensity and shape before fabrication of TiO layers. The hexagonal pattern in the Au layer (Fig. 2(a), bottom) was fabricated by electron-beam lithography with ZEP-420 resist and subsequent Ar-ion milling in an area of 27 × 27 $\mu m^2$. We prepared samples with different periodicities of the hexagonal arrays to study the effect on vertical photoluminescence (PL) extraction efficiency. The hole radius in hexagonal arrays is chosen to keep the filling factor at 11 %. The filling factor was chosen the smallest to extract enough PL for decay-time measurement, in order to compare the measurement with theoretical analysis on unpatterned structures. Subsequent deposition of Si on top of the patterned Au layer leads to Si-filled holes in the Au plasmonic crystal. For comparison, we also prepared InP-TiO-Au-TiO structures and samples without the hexagonal arrays.

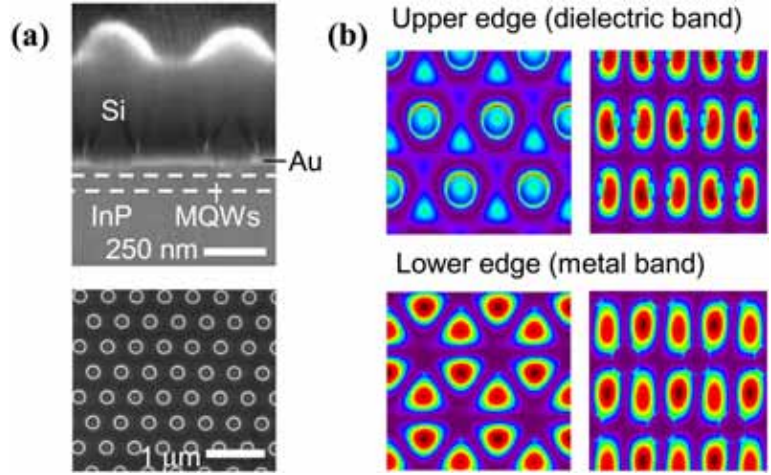

Fig. 2. Scanning electron microscope images of fabricated plasmonic crystal patterns and the corresponding field patterns. a) SEM images of the cross-section of fabricated InP-TiO-Au-TiO-Si structure (top) with a hexagonally patterned Au layer (bottom) with periodicity $a$ = 480 nm and hole radius $r/a$ = 0.18. The bumps on the surface of the Si-waveguide result from the Si deposition on top of the hexagonal pattern in the Au layer. b) FDTD simulations of the lateral field distributions of SPP modes in hexagonal plasmonic crystals in case of the symmetric index on top and on bottom; simulated structure consists of 20nm Au layer sandwiched by InP on both sides, and crystal periodicity $a$ and hole radius $r$ are defined as: $a$ = 450 nm and $r/a$ = 0.2. The modes shown correspond to the anti-symmetric SPP band folder back to the $\Gamma$-point (vertical emission) by the crystal. The left and right panels show the monopole and dipole components of the field belonging to the same band edges, respectively. Eigen fields at the band edges ($\Gamma$-point) could not be separated because of close overlap of their dispersion branches.

*3.2 Photoluminescence measurement*

We first consider the coupling between the MQWs and the plasmonic crystal by measuring PL intensity into crystals with varying periodicities. In the PL measurements, we optically excited MQWs from the Au layer side by a normally incident laser beam focused with a 100x objective lens with NA=0.65. The excitation wavelength was 980 nm. The PL from the sample was collected with the same objective lens and detected with a spectrum analyzer or a streak camera. We observed TM-polarized photoluminescence (PL) due to the light-hole (LH) transition near 1,350 nm for the bare MQW wafer; the primarily PL due to the heavy-hole (HH) transition was completely suppressed at room temperature. Figures 3(a) and 3(b) show the spectra for InP-TiO-Au-TiO and InP-TiO-Au-TiO-Si structures with and without hexagonal arrays of various lattice constants ($a$). The blue and red arrows in Figs. 3(a) and 3(b) indicate the theoretically estimated frequencies of the symmetric and WC-anti-symmetric modes with $k = 4\pi/\sqrt{3}a$ in the structure without hexagonal pattern. Those points are of particular interest because they are folded back to the $\Gamma$ point in plasmonic crystal, corresponding to vertical emission [15]. In the PL spectra for InP-TiO-Au-TiO shown in Fig. 3(a), we observed an enhanced PL signal which shifts toward shorter frequencies with an increase in $a$. These PL enhanced spectra have a maximum when the frequencies indicated by the blue arrows cross the spectrum of the MQW gain around 1,350 nm, which proves that these PL signals correspond to the symmetric SPP modes (see Fig. 1) vertically scattered by the hexagonal array with the same reciprocal vector as the wave vector of SPPs ($\Gamma$ point). A slight mismatch between the PL maxima and the positions of blue arrows is due to the InP oxide created during the Ti oxidation process. The background signals independent of $a$, which were extremely small in the PL spectrum for the structure without hexagonal pattern

(red spectrum), arise from uncoupled excitons located beneath the holes in plasmonic crystal. In the PL spectra for InP-TiO-Au-TiO-Si, we expect to see two kinds of enhanced PL signals corresponding to the two SPP modes: symmetric and WC-anti-symmetric (see Fig. 1). These are visible in the spectra with $a$ = 380 - 500 nm in Fig. 3(b). Comparing Figs. 3(a) and 3(b), we note that the PL signals corresponding to symmetric modes are shifted toward lower frequencies because of approximate index matching with the extra Si layer, as expected from Fig. 1(a) [10]. The intensities of the signals corresponding to the symmetric modes were smaller than those corresponding to WC-anti-symmetric modes because of their larger metal absorption loss [4,5]. The PL peak observed in the spectrum for $a$ = 540 nm (top plot in Fig. 3(b)) coincides with the light line for TiO layers ($n$ = 2.5), and we expect that the structural periodicity of the Si waveguide induced by the hexagonal arrays probably scatters the light traveling through the TiO layers.

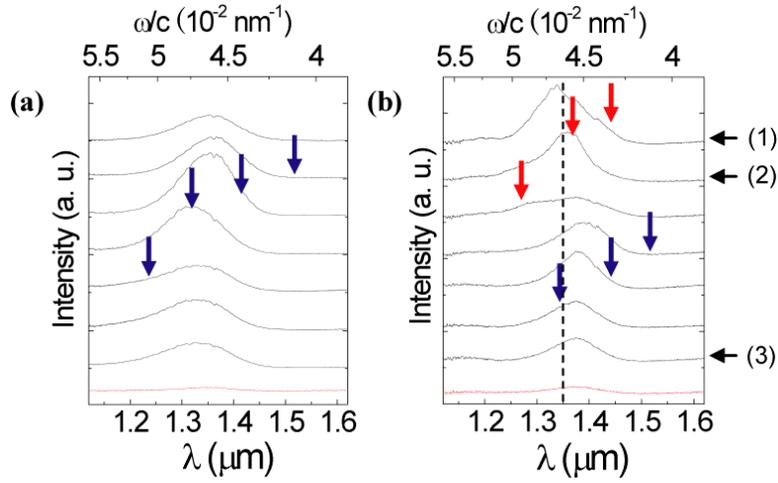

Fig. 3. Photoluminescence measurements in proposed plasmonic crystals. a) spectra for InP-TiO-Au-TiO hexagonal plasmonic crystals. b) spectra for InP-TiO-Au-TiO-Si hexagonal plasmonic crystals. Lattice constants ($a$) of the crystals are 540, 500, 460, 420, 380, 340, 300 nm respectively from the top. The red plots (bottom) show the spectra for the samples without hexagonal patterns. Vertical dashed line indicates the wavelength of 1,350m for which the time-resolved PL measurements in Fig. 4 are taken. Arrows indicate the theoretically estimated frequencies of symmetric (blue) and WC-anti-symmetric (red) modes with $k = 4\pi/\sqrt{3}a$ in the structure without hexagonal pattern.

## 3.3 Time-resolved (decay-time) measurement

The coupling efficiency of PL into plasmon modes can be increased substantially in our structure by enhancing the spontaneous emission rate via the Purcell effect. In Figs. 4(a) and 4(b), we directly measure this Purcell enhancement with time-resolved PL measurements. Figure 4(a) shows the initial PL decay rates defined as normalized time derivatives of the initial PL decay $(\delta PL / \delta t)_{t=0} / PL_{t=0}$ measured at the wavelength of 1,350 nm for spectra (1), (2) and (3) shown in Fig. 3(b) and for a bare MQW wafer, with different pumping powers $I$. Figure 4(b) shows the PL decay curves for the spectrum (2) (top plot) and for a bare MQW wafer (bottom plot), with the pumping power of $I$ = 6 mW. To experimentally evaluate the Purcell enhancement factor from the time-resolved PL measurements shown in

Fig. 4, we considered the Boltzmann approximation for decay rate of the carrier density $N$. In this approximation, the time derivative of $N$ in non-doped semiconductors is expressed as a polynomial of $N$:

$$dN/dt = -AN - BN^2 - CN^3, \qquad (1)$$

where the coefficients $A$, $B$, and $C$ correspond to surface- or defect-related recombination, bimolecular recombination, and an Auger process, respectively [17]. Among the three recombination processes in Eq. (1), only the second term contributes to PL radiation: $PL \propto BN^2$, and thus $(dPL/dt)_{t=0}/PL_{t=0}$ is expressed by the following equation:

$$(dPL/dt)_{t=0}/PL_{t=0} = -2A - 2BN_0 - 2CN_0^2. \qquad (2)$$

The values of $I$ for the measurement were much lower than the saturation power for excitation of MQWs, and thus the initial carrier density $N_0$ could be considered proportional to $I$ ($N_0 \cong \eta I$ where $\eta$ is the excitation efficiency). Hence, Eq. (2) can be also rewritten as a polynomial of $I$. Since the observed values of $(\delta PL/\delta t)_{t=0}/PL_{t=0}$ change linearly with $I$ in Fig. 4(a), we could neglect the Auger process represented by the third term in Eq. (2) under our measurement condition. In this case, from the slope and offset of results in Fig. 4(a), we estimate $A^b = 0.54 \times 10^9\ s^{-1}$, $A^m = (0.63 \pm 0.03) \times 10^9\ s^{-1}$, and $B^m \eta^m / B^b \eta^b = 1.5 \pm 0.2$, where superscripts $b$ and $m$ indicate the values for a bare MQW wafer and the InP-TiO-Au-TiO-Si structures, respectively. The large values of $A$ could be due to the deformation caused by the large tensile strain applied in MQWs. The slopes of the fitting lines for spectra (1) (2) and (3) are almost the same within plotting errors. We can expect the position-dependence of the radiative decay of the excitons within a unit cell of our plasmonic crystal from the field patterns in Fig. 2(b) [18]. However, the thermal diffusion length of carriers in GaInAsP during the typical decay time of 100 ps is estimated on the order of 10 μm, much longer than the radii of the holes in the hexagonal pattern [19]. Hence the position-dependence of the radiative decay of the excitons within a unit cell of our plasmonic crystal is not observable in this measurement. We estimated the value of $B^m \eta^m / B^b \eta^b \approx 1.5$ by fitting the decay curves shown in Fig. 4(b) with the PL decay function obtained by solving Eq. (1) and assuming $C = 0$:

$$PL(t) \propto B\{N(t)\}^2 \propto \left\{\frac{e^{-At}}{A + B\eta I(1 - e^{-At})}\right\}^2, \qquad (3)$$

For this fit, the values of $A^b$ and $A^m$ were estimated from the offsets of the plots at $I = 0$ in Fig. 4(a). We extract the Purcell enhancement from $B^m \eta^m / B^b \eta^b$. $\eta^m / \eta^b$ can be replaced by the ratio of the transmissions of the pump beams through plasmonic crystal and bare MQW wafer, which is estimated to be 0.45 by spatial averaging (it was confirmed by FDTD simulation that there were no SPP modes coupled to the incident pumping beam, enabling the resonant enhanced transmission [20]). Therefore, the Purcell enhancement factor with non-radiative decay ignored, which should be identical to the ratio of $B^m / B^b$, is estimated to be 2.9-3.7.

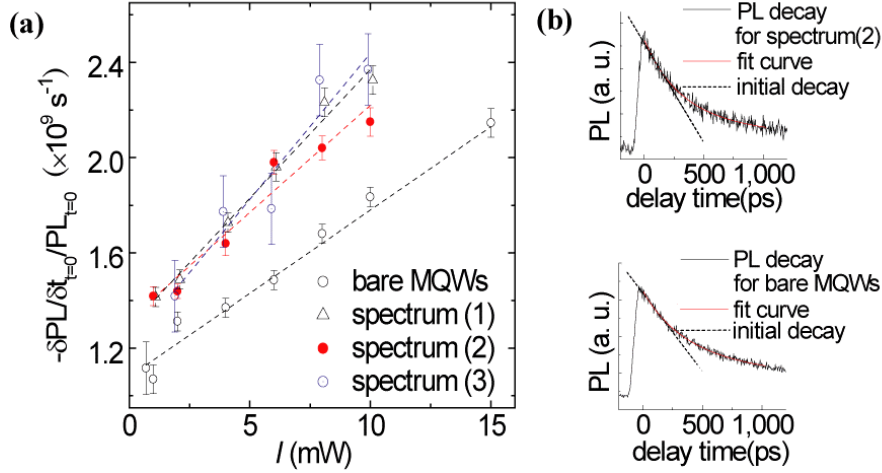

Fig. 4. Time-resolved photoluminescence measurements. a) Initial slope of the PL decay $(\delta PL/\delta t)_{t=0}/PL_{t=0}$ plotted as a function of the pump power $I$, for a bare MQW wafer and for spectra (1), (2) and (3) in Fig. 3(b) at the denoted wavelength of 1,350 nm. Dashed lines are fits to the experimental data. b) PL decay curves for a bare MQW wafer (bottom plot) and for spectrum (2) (top plot) with $I$ = 6 mW. The excitation wavelengths was 750 nm.

## 4. Theoretical analysis for Purcell enhancement factor

To theoretically estimate the Purcell enhancement factor, we modeled the InP-TiO-Au-TiO-Si structure without a hexagonal pattern. Since the excitons lying beneath the Au layer are coupled to both SPP and non-SPP modes, the total Purcell enhancement factor is represented by summing all of these contributions:

$$F = F_{non-SPP} + F_{WC-antisym-SPP} + F_{sym-SPP}. \qquad (4)$$

From FDTD simulations, the coupling of excitons to the parallel component of the SPP field is negligible because $E_{//}^2/E_{\perp}^2 < 0.1$ at 1,350 nm and at the MQW position. With this in mind, we obtained the form of Purcell factor for both WC-anti-symmetric and symmetric modes in an unpatterned structure [9]:

$$F_{SPP} = \frac{2}{3} \times \frac{3\pi c^3 k E_{\perp}^2(a)}{2n\omega^2 \int_{-\infty}^{\infty}[\partial(\omega\varepsilon)/\partial\omega]E^2(z)dz}\frac{dk}{d\omega}, \qquad (5)$$

where the non-radiative decay is ignored. The factor 2/3 in Eq. (5) comes from the ratio of the coupling strengths of the electric dipole parallel to $E_{\perp}$ and $E_{//}$ for LH transition [15,17], which constitutes the dominant radiative process in our MQW structure. In Eq. (5), we also assumed that the coupling strengths for $E_{\perp}$ are independent of the propagation directions of SPPs. By using the dispersion relations and the spatial distribution of the photon energy densities shown in Fig. 1, $F_{WC-antisym-SPP}$ and $F_{sym-SPP}$ were estimated to be 0.8 and 2.8, respectively. A smaller group velocity and strong field confinement (which can be seen in Fig. 1) for symmetric modes lead to $F_{sym-SPP}$ larger than $F_{WC-antisym-SPP}$. With $F_{non-SPP} \approx 1$, Eq. (4) gives an overall Purcell factor of $F \approx 4.6$.

Our experimentally estimated value of $F$ falls within 75-80% of this theoretical estimate. One contributor to the reduction in $F$ is InP oxide created on the InP surface during fabrication, which lowers the refractive index and shifts the mode toward these layers and away from the MQW position (such a mode shift is also visible for TiO layers, as shown in Fig. 1(c)). We speculate that another contributor is the presence of uncoupled excitons located under dielectric holes in the hexagonal array due to large leakage losses in the dielectric band, which redirect part of the PL emission into the non-SPP modes (background PL).

## 5. Conclusion

In conclusion, we proposed and tested a symmetric plasmonic crystal that is highly tolerant to fabrication imperfections, while supporting an anti-symmetric-like mode which significantly suppresses ohmic losses and shifting a symmetric mode toward lower frequencies. Although the field pattern is not precisely anti-symmetric, the small photon energy density within the metallic region still provides small absorption loss to the WC-anti-symmetric modes [4]. By FDTD simulation, we estimate the quality factor due to the metal absorption at $Q_{ab} \approx 140$ for the WC-anti-symmetric mode at 1,350 nm, which is more than 50 % of the corresponding Q-factor in a completely symmetric InP-Au-InP structure. A plasmonic crystal integrated into the structure redistributes the in-plane field patterns into metal and dielectric bands, and allows us to achieve high Purcell enhancement for the excitons beneath metal. The structure offers enhanced spontaneous emission rate of up to 3.7 times. Theoretically we find that 80 % of the enhanced PL is emitted into SPP modes, and 17 % can be extracted into WC-anti-symmetric modes, which is more than 5 times higher than the extraction efficiency for the unpatterned semiconductor wafer (3 %). Redirection of the emission into symmetric modes with lower group velocities is one of the causes to decrease the extraction into low-loss WC-anti-symmetric modes. Since symmetric modes tend to have larger photon energy density on the high-index side, as opposed to anti-symmetric modes, the replacement of Si layer with a little higher index material could shift those modes to the opposite side of MQW, possibly improving the extraction into WC-anti-symmetric modes. Another possibility to improve the extraction into WC-anti-symmetric modes is to use a thicker Au layer. In this case, symmetric modes have higher group velocities, consequently decreasing the redirection into those modes. Although use of higher index material instead of Si or thickening Au layer may disgrace the low-loss in WC-anti-symmetric modes, the optimization between the extraction efficiency and ohmic loss in WC-anti-symmetric modes should be important for plasmonic lasing or direct couplers between excitons and plasmonic waveguides. The proposed and demonstrated plasmonic crystal marks an important step towards SPP laser cavities [11-13], the development of highly integrated photo-electrical devices [7,8], or miniaturized biosensors [6].

## Acknowledgements

This work has been supported by Production Engineering Research Laboratory of Canon Inc., the MARCO IFC, and the NSF Grant No. CCF-0507295.